\documentclass[12pt]{iopart}
\usepackage{iopams}
\usepackage{epsfig}   
\usepackage{graphics}
\begin{document}

\title[No Evidence for Dark Energy Metamorphosis ?]{No Evidence for Dark
  Energy Metamorphosis ?}

\author{J. J\"onsson, A. Goobar, R. Amanullah and L. Bergstr\"om}

\address{Department of Physics, Stockholm University, Albanova
         University Center \\
         S--106 91 Stockholm, Sweden}
\begin{abstract}
Recent attempts to fit Type Ia supernova data by 
modeling the dark energy density as a truncated
Taylor series have suggested the possibility of {\em metamorphosis}, i.e., 
a rapidly evolving equation of state parameter, $w_{DE}(z)$. However,
we show that fits
using that parametrization have significant problems.  
Evolution of $w_{DE}(z)$
is both favoured and in some sense forced, and the
equation of state parameter blows up or diverges in large regions of
the parameter space. 
To further elucidate these problems we have simulated sets of supernova data
in a $\Lambda$--universe to show that the suggested
``evidence'' for metamorphosis is also common for $w_{DE}=-1$.    
\end{abstract}

\eads{\mailto{jacke@physto.se}, \mailto{ariel@physto.se}, 
\mailto{rahman@physto.se}, \mailto{lbe@physto.se}}




\section{Introduction}
Revealing the true nature of dark energy (DE) has become one of the
most important tasks in cosmology.   
Considering the plethora of DE models proposed in the literature, a
model independent reconstruction of DE 
would be an appealing alternative to testing all models separately. 
In two
recent papers~\cite{alam:dec03,alam:mar04} attempts were made by Alam
\etal to reconstruct the dark energy equation of state parameter
$w_{DE}(z)$ in a model independent manner, using the latest supernova
data~\cite{hzt:tonry,hzt:barris,hzt:riess,scp:knop}. In these two
papers a truncated Taylor series was used to model the dark energy
density $\rho_{DE}(z)$.  The results indicate an evolution of
$w_{DE}(z)$, a behaviour they call metamorphosis. From the reported
analysis, it would seem that this is a significant effect 
prompting for ``exotic'' models for the DE.

Since other parameterizations~\cite{hzt:tonry,hzt:riess,scp:knop}
suggest that the Type Ia supernova data
collected so far are consistent with the simplest DE model of all,
that of a cosmological constant $\Lambda$ ($w_{DE}=-1$), it is important
to investigate how such different conclusions can be 
reached starting from the same sets of
data.  In this paper we argue that the method of model independent
reconstruction proposed by Alam \etal suffers from a number of serious
shortcomings. For alternative methods of DE reconstruction see e.g. 
references~\cite{wang1,wang2,wang3,huterer,daly}.

A fundamental requirement on 
a model independent
reconstruction of $w_{DE}(z)$ must be that 
all DE alternatives are treated on an equal footing.
Although the method of Alam \etal at first sight
seems to be capable of an exact reproduction of the equation of state
parameter of the cosmological constant, it actually favours evolving
$w_{DE}(z)$. The confidence contours, describing the level to which
the reconstructed $w_{DE}(z)$ is known, exhibit two related problems. 
First, contours enclosing regions of high confidence level (CL) 
and high redshift
tend to diverge. Second, by construction, the 
low level CL regions shrink for high redshifts.

\section{Confronting supernova data with the model}

In this section we present the basic formulae needed for confronting the model
of Alam \etal with supernova data.
The ansatz for the dark energy density proposed by Alam \etal is a
truncated Taylor series
\begin{equation}
\label{eq:ansatz}
\rho_{DE}=A_{0}+A_{1}x+A_{2}x^{2},
\end{equation}
where $x=1+z$. In the following, we will assume that the universe is
flat and dominated by matter and DE. The Hubble parameter in
such a universe, with the DE parametrized as in equation~(\ref{eq:ansatz}), 
is hence given by
\begin{equation}
\label{eq:hubble}  
H(x)=H_{0}(\Omega_{M}x^{3}+A_{0}+A_{1}x+A_{2}x^{2})^{1/2}.
\end{equation}
If the matter density is assumed to be known, only two of the
parameters describing the DE are independent,
$A_{1}$ and $A_{2}$. The third parameter is given by the normalization
condition, $1=\Omega_{M}+A_{0}+A_{1}+A_{2}$.

The equation of state parameter $w(z)=p(z)/\rho(z)$ relates the pressure to the
energy density and can be used to characterize an energy species. The
dark energy equation of state parameter $w_{DE}(z)$ can be calculated
given the Hubble
parameter and the matter density~\cite{saini}
\begin{equation}
\label{eq:w(H)}
w_{DE}(x)=\frac{(2x/3)d(\ln H)/dx-1}{1-(H_{0}/H)^{2}\Omega_{M}x^{3}}.
\end{equation}
By using the parametrization in equation~(\ref{eq:hubble}), 
$w_{DE}(x)$ can be expressed in terms of the parameters $A_{0}$,
$A_{1}$ and $A_{2}$ 
\begin{equation}
\label{eq:de_eos}
w_{DE}(x)=-\frac{A_{0}+\frac{2}{3}A_{1}x+\frac{1}{3}A_{2}x^{2}}{A_{0}+A_{1}x
+A_{2}x^{2}}.
\label{eq:de_para}
\end{equation}
Using the derivative of equation~(\ref{eq:hubble}) in the
derivation  of equation~(\ref{eq:de_eos}), is however a questionable
step. If the
parametrization of $H(x)$ is
incorrect, $dH(x)/dx$ is likely to be even more erroneous. 
The parametrization~(\ref{eq:de_eos}) is capable of reproducing the 
equation of state parameter for the cosmological constant 
($w_{DE}=-1$, $A_{1}=A_{2}=0$), and
topological defects in the form of cosmic strings
($w_{DE}=-\case13$, $A_{0}=A_{1}=0$) and domain walls     
($w_{DE}=-\case23$, $A_{0}=A_{2}=0$).

Models of DE are tested against observations 
by deriving distances to Type Ia supernovae through their measured
brightness. Cosmic distances are related through
the $H_{0}$--independent luminosity distance
$d_{L}'$ to the brightness of a supernova measured in magnitudes
\begin{equation}
\label{eq:mag}
m=5\log d_{L}'+\mathcal{M},
\end{equation}
where $\mathcal{M}=M+25-5\log H_{0}$ will be treated as a ``nuisance''
parameter in the fitting procedure and the absolute
magnitude of a supernova is denoted by $M$. The expression for the
$H_{0}$--independent
luminosity distance in a flat universe is
\begin{equation}
\label{eq:dL}
d_{L}'(z)=(1+z) \int _{1}^{1+z} \frac{H_{0}}{H(x)} dx.
\end{equation}
Observed and theoretically predicted magnitudes can thus be
compared by means of the equations (\ref{eq:mag}) and (\ref{eq:dL}) and the
model of the DE enters via equation (\ref{eq:hubble}).
However, since the integrand in~(\ref{eq:dL}) has to be real, i.e. 
$\Omega_{M}x^{3}+A_{0}+A_{1}x+A_{2}x^{2} \geq 0$, only a  part of
the $A_1-A_2$ parameter plane is allowed. As long as this condition is
fulfilled, the dark energy density can be negative.

\section{Reconstructing the dark energy equation of state parameter}
\label{sec:reconstruction}

In this section we reconstruct the dark energy
equation of state parameter by the method proposed by Alam \etal. 

We have used two sets of supernova data, which will be referred to as
the \emph{small} and the \emph{large} data set. The \emph{small} data set is a
combination of supernovae from Tonry \etal~\cite{hzt:tonry} and
Barris \etal~\cite{hzt:barris}, and it is the
same set as Alam \etal used in reference~\cite{alam:dec03}. The $172$
supernovae in the \emph{small} set have low extinction $A_{V}<0.5$ and 
$z>0.01$.
The \emph{large} data set is an extension of the \emph{small} set with 
$16$ additional high redshift supernovae from Riess \etal~\cite{hzt:riess}.
 
A maximum likelihood analysis of the data sets was used to estimate
the best fit parameters of the dark energy model. The following
negative log--likelihood function was used
\begin{equation}
\label{eq:L}
\mathcal{L}=-\ln \sum_{i=1}^{N}\frac{1}{2}\left(
\frac{m_{i}-\left(5\log
  d_{L}'(\Omega_{M},A_{1},A_{2};z_{i})+\mathcal{M}\right)}
  {\sigma_{i}}\right)^{2},
\end{equation}
where $m_{i}$ and $\sigma_{i}$ are observed magnitude and
magnitude error
of a supernova at redshift $z_{i}$. The parameter $\mathcal{M}$, related to
the Hubble constant and the absolute supernova magnitude, was
treated as a ``nuisance'' parameter in the fitting procedure described below.

First, the negative log--likelihood function in equation (\ref{eq:L})
was computed on a cubic
lattice in the parameter space spanned by  $\Omega_{M}$,
$A_{1}$ and $A_{2}$. Thereafter a Gaussian
prior $\Omega_{M}=0.29 \pm 0.07$, containing
information on the matter density, was applied to
 $\mathcal{L}$\footnote{Note that the specific
choice of prior of $\Omega_{M}$ is not critical for our argument.}. 
The $A_1-A_2$ plane
containing the minimum of the negative log--likelihood function was then
identified. Finally, the best fit values and the CL contours in
this plane were translated into $w_{DE}(z)$ through equation
(\ref{eq:de_para}).

\begin{figure}
\begin{center}
\resizebox{1.0\textwidth}{!}{\includegraphics{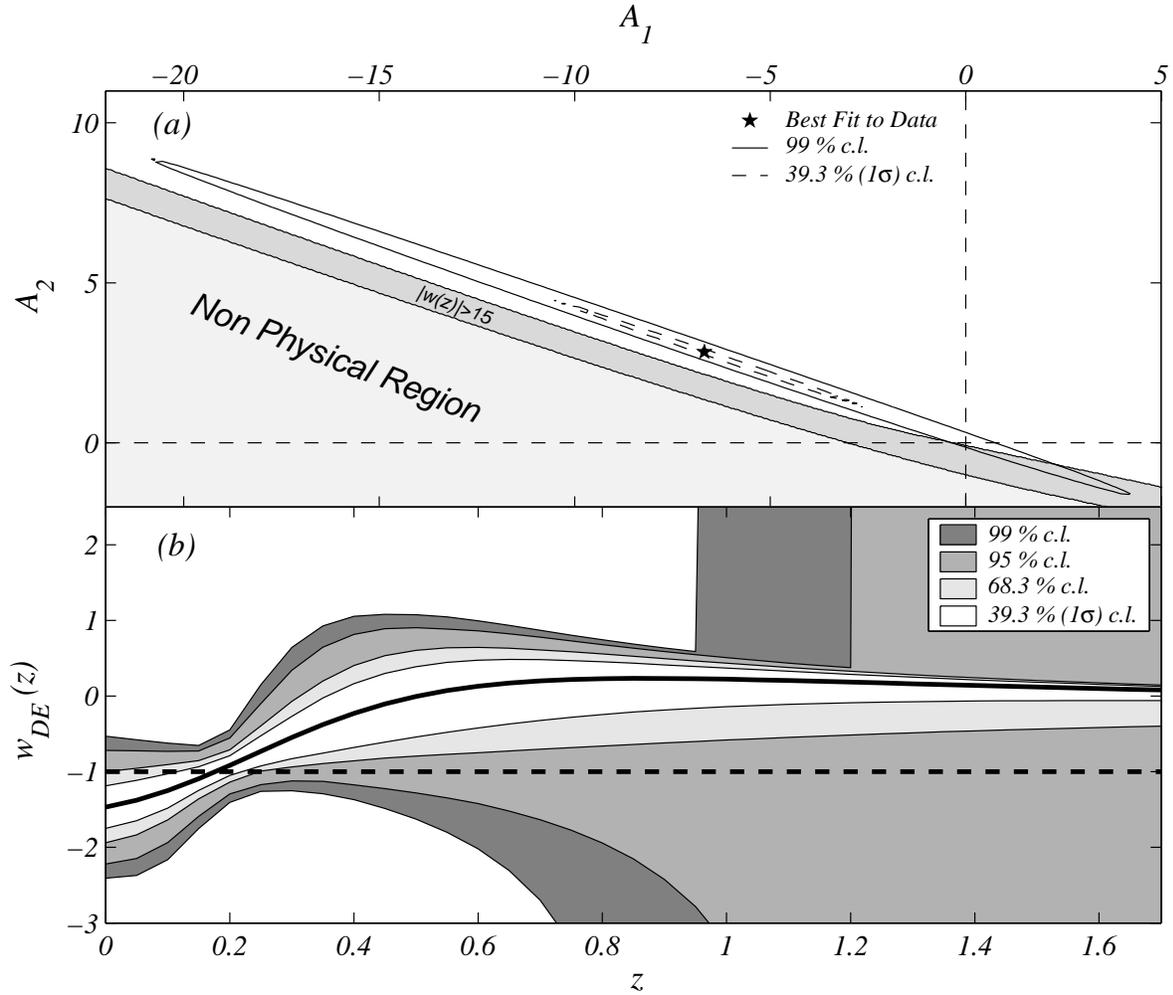}}
\caption{\label{fig:hzt_w} The results of the reconstruction of DE from
the \emph{small} data set, consisting of supernovae from Tonry \etal and
Barris \etal. 
  Panel (a) shows
  the $A_1 - A_2$ parameter plane, where the
  best fit value to the data ($A_{1}=-6.68$ and $A_{2}=2.84$) is
  indicated by a star. The solid and dashed contours indicate the 
  $99\%$ and 1$\sigma$ confidence levels (CL) respectively. 
  In panel (b) the reconstructed equation of state parameter
  corresponding to the best fit (solid line) and various confidence 
  regions are drawn. The obtained equation of state
  parameter is consistent with a cosmological constant at the $95\%$
  CL, but not at the $68\%$ CL. 
  Note that Alam {\it et al} report their confidence regions as
  1$\sigma$ levels, i.e. as $\mathcal{L}_{max} - {1 \over 2}$ in their
  maximum likelihood analysis. As two parameters are jointly fitted this
  corresponds only to a 39.3 \% CL.  The shown discontinuities in the
  $95\%$ and $99\%$ CL regions correspond to the contours crossing the
  dark shaded region in the $A_1-A_2$ plane in panel (a), where $w_{DE}(z)$
  grows very rapidly.}
\end{center}
\end{figure}

The best fit parameters to the \emph{small} data set 
($A_{1}=-6.68$ and $A_{2}=2.84$) and the $99\%$ and 1$\sigma$ CL
contours in the parameter plane are shown in panel (a) in
figure~\ref{fig:hzt_w}. The 1$\sigma$ CL contour is drawn in order to
facilitate comparison with reference~\cite{alam:dec03} and~\cite{alam:mar04}.
Note that Alam {\it et al} report their confidence regions as
1$\sigma$ levels, i.e. as $\mathcal{L}_{max} - {1 \over 2}$ in their
maximum likelihood analysis. As two parameters are jointly fitted this
corresponds only to a $39.3\%$ CL.
The reconstructed history of the dark energy equation of state
parameter, shown in panel (b) in figure~\ref{fig:hzt_w},
is clearly consistent with a cosmological constant at the $95\%$, but
not at the $68\%$ CL. There are however some peculiarities of
the reconstructed $w_{DE}(z)$ that should be noted. 
The $95\%$ and $99\%$ confidence regions drawn in
figure~\ref{fig:hzt_w} diverge at large redshifts. This corresponds to
the contour regions crossing the dark shaded region in the $A_1-A_2$ plane, 
where $w_{DE}(z)$ grows very rapidly. The $68\%$ and 1$\sigma$ confidence
regions, on the other hand, converge at large redshift.
Moreover, there is a waist at $z \approx 0.2$ where all contours
are squeezed together.  
These peculiarities indicate that this reconstruction is
in fact not model independent, as will be further explained in the
next section. 

\begin{figure}
\begin{center}
\resizebox{1.0\textwidth}{!}{\includegraphics{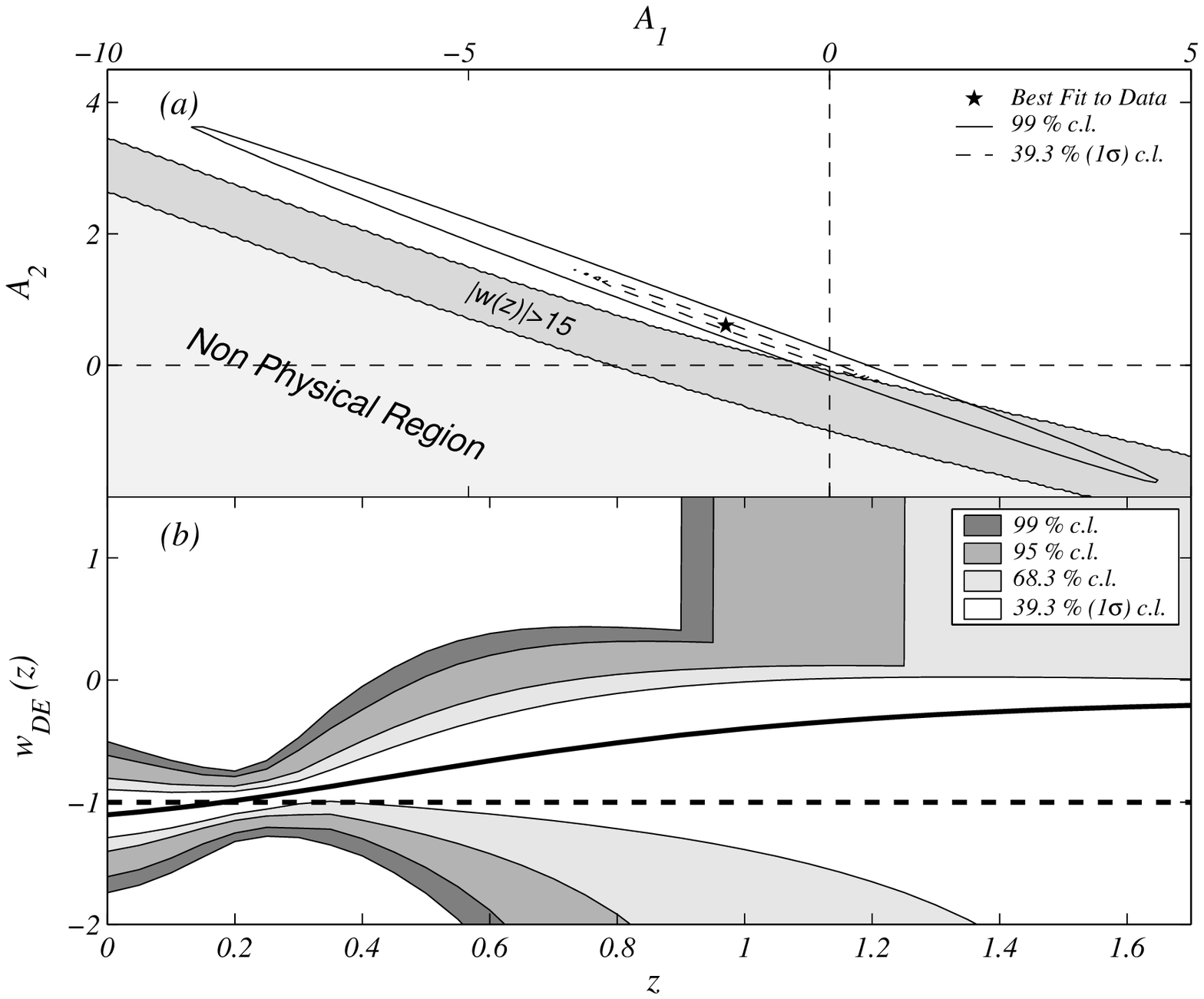}}
\caption{\label{fig:hzt_16hz} The results of the reconstruction of DE from
the \emph{large} data set, consisting of the \emph{small} data set and $16$
additional high--z supernovae from Riess \etal.
In panel (a) the best fit values to this
data set ($A_{1}=-1.44$ and $A_{2}=0.61$) is indicated by a star. The solid and
dashed contours indicate the $99\%$ and 1$\sigma$ confidence
levels (CL) contours respectively. Both CL contours
cross the region in the parameter plane where $w_{DE}(z)$ blows up or
diverges (dark gray). 
The equation of state obtained for this data set, shown in panel (b), is
consistent with the cosmological constant
  not only at the $95\%$ CL, but also at the $68\%$ CL.}
\end{center}
\end{figure}

Figure~\ref{fig:hzt_16hz} shows $w_{DE}(z)$
obtained from the \emph{large} data set with $16$ additional supernovae.
This equation of state parameter is consistent with the cosmological constant
even at the $68\%$ CL. However, the confidence regions 
still exhibit the same behaviour as the ones obtained for the \emph{small}
data set.

\section{\label{sec:analysis} Analysis of the dark energy equation of 
state parameter}
The unusual behaviour of the results in figure~\ref{fig:hzt_w} 
and~\ref{fig:hzt_16hz} appear since
the dark energy equation of state parametrization~(\ref{eq:de_eos}) has a few
peculiarities, which we discuss in this section.

\subsection{Diverging equation of state parameter}
The dark energy equation of state parameter will obviously diverge whenever the
denominator in equation~(\ref{eq:de_eos}) vanishes. Regions in the parameter
plane where $w_{DE}(z)$ blows up ($|w_{DE}|>15$) or diverges somewhere in
the redshift interval $0 \leq z \leq 1.8$ are shown in dark gray in
e.g. panel (a) in figure~\ref{fig:hzt_w}. The point $A_{1}=A_{2}=0$, 
corresponding to the cosmological constant, lies very close to this
region. If the best fit to data is near this point, the error in
$w_{DE}(z)$ will blow up.
Diverging equation of state parameter is the reason why the $95\%$
confidence region in e.g. figure~\ref{fig:hzt_w}(b) blows up. 

\subsection{High redshift limits and shrinking confidence regions} 
From equation~(\ref{eq:de_eos}) one can easily see
that the parametrized equation of state depends on the order of the 
Taylor expansion
at the high redshift limit. If, for instance, only the first order is
considered, $w_{DE} \rightarrow -\case23$ for high redshifts, while
for the second order $w_{DE} \rightarrow -\case13$.
Since the proposed ansatz~(\ref{eq:ansatz}) is a truncated Taylor
series we cannot expect equation~(\ref{eq:de_eos}) to be
valid at very high
redshifts. However, the effects of these high redshift limits are noticeable
already at the moderate redshifts where we try to model DE.
These limits are responsible for the shrinkage of the non--diverging
confidence regions at high redshift (e.g., see the $68\%$ confidence
region in figure~\ref{fig:hzt_w}(b)). 

\subsection{Stability of the dark energy equation of state parameter} 
The dark energy equation of state parameter described by 
equation~(\ref{eq:de_eos}) can
be thought of as a family of curves, where each curve is parametrized by
$\Omega_{M}$, $A_{1}$ and $A_{2}$. In this section we 
study the stability of these curves to perturbations in the
parameters $A_{1}$ and $A_{2}$. Moreover, we will show
that some curves in this family become more stable with
redshift, with respect to these perturbations, while
others become more unstable.
The change
$\delta w_{DE}$ in a curve, described by the parameters $A_{1}^{0}$ and
$A_{2}^{0}$, due to small perturbations $\delta A_{1}$ and $\delta A_{2}$
is given by
\begin{equation}
\delta \!  w_{DE} \approx \frac{\partial w_{DE}}{\partial A_{1}}      
(A_{1}^{0},A_{2}^{0}) 
\delta \! A_{1}+\frac{\partial w_{DE}}{\partial A_{2}}  (A_{1}^{0},A_{2}^{0}) 
\delta \! A_{2}.
\end{equation}  
The curve that corresponds to the cosmological constant
($A_{1}=A_{2}=0$) becomes increasingly unstable due to perturbations as
redshift increases,
\begin{equation}
\delta \! w_{\Lambda} \approx \frac{x}{3(1-\Omega_{M})} \delta \! A_{1} +
\frac{2x^{2}}{3(1-\Omega_{M})} \delta \! A_{2}.
\end{equation}
A curve that starts close to the line $w_{DE}(x)=-1$ will thus
departure from this line with increasing redshift. 
Equation~(\ref{eq:de_eos}) can also 
make an exact reproduction of two types of topological
defects. The equation of state of domain walls
($A_{0}=A_{2}=0$) grows stable to 
perturbations in $A_{1}$ with $z$, but not to perturbations in $A_{2}$
\begin{equation}
\delta \! w_{DE} \approx \frac{1}{3(1-\Omega_{M})x}\delta \! A_{1}+ 
\frac{1}{3(1-\Omega_{M})} \left( x+\frac{1}{x} \right) \delta \! A_{2}.
\end{equation}
The curve corresponding to cosmic strings ($A_{0}=A_{1}=0$) is stable
to perturbations
in both $A_{1}$ and $A_{2}$ at high redshifts
\begin{equation}
\delta \! w_{DE} \approx \frac{2}{3(1-\Omega_{M})} \left(\frac{1}{x^{2}}-
\frac{1}{2x} \right) \delta \! A_{1}+ 
\frac{2}{3(1-\Omega_{M})x^{2}} \delta \! A_{2}.
\end{equation}
The curves corresponding to topological defects are thus very
special. All other curves (except the one corresponding to the
cosmological constant) approach one of these two curves as redshift increases.
The effect of perturbations in $A_1$ and $A_2$ depends on the value of
these parameters. This is in contrast to perturbations of linear
parameterizations of $w_{DE}(z)$, e.g. $w_{DE}(z)=w_0+w_1z$, which are 
independent of parameter values.  
From the above discussion
about the stability of $w_{DE}(x)$ with redshift, we
see that the parameterization in equation~(\ref{eq:de_eos}) treats the
cosmological constant unfairly in favour of other DE alternatives.

\subsection{Forced evolution of the dark energy equation of state parameter}
\label{sec:force}
The parametrized equation of state parameter (\ref{eq:de_eos}) will
not only favour,
but also ``force'' evolution. All curves, describing an evolving
equation of state parameter, will cross at a
certain redshift. 
At this redshift $z_{*}=x_{*}-1$, where $x_{*}=-A_{1}/2A_{2}$, the numerator
and denominator are equal and the dark energy equation of state
assumes the value $w_{DE}(z_{*})=-1$.
The parameters $A_{1}$ and $A_{2}$, obtained by fitting supernova data to the
ansatz (\ref{eq:ansatz}), are highly correlated and the ratio $A_{1}/A_{2}$
is thus nearly constant. The value of this ratio depends on the
slope of the CL contours in the $A_1-A_2$ plane
which in turn is correlated with the assumed prior on the matter density 
$\Omega_{M}$.
This implies that the location of the point
$x_{*}=-A_{1}/2A_{2}$ is
more or less independent of the parameters $A_{1}$ and $A_{2}$ for a
given $\Omega_{M}$. The parametrization~(\ref{eq:de_eos}) of the equation
of state parameter thus favours DE
evolution. If the reconstructed
equation of state at present is less (larger) than
minus one it must have been larger (less) than minus one in the past.
Any significant departure from the line $w_{DE}(x)=-1$ will hence
force the reconstructed
equation of state parameter to cross that line. Since most curves will pass
through this point, the family of curves will have a waist centered at
$x_{*}=-1/2k$, where $k$ denotes the slope of the CL
contours in the $A_1-A_2$ plane.
The transition redshift in e.g.
figure~\ref{fig:hzt_w} is $z_{*} \approx 0.2$ which corresponds to
$k = -0.42$, in good agreement with the slopes of the ellipses in
panel (a).

\section{Simulations}
To test the reliability of this way of reconstructing the dark energy 
we tried to recover a fiducial cosmology from 
simulated data sets generated using
SNOC~\cite{snoc}, a Monte--Carlo simulation package for high--z
supernova observations \footnote{The SNOC code is available at
http://www.physto.se/\~{}ariel/snoc.}. 
We generated $500$ data sets, each one consisting of $172$
supernovae with the same distribution of redshifts and magnitude
errors as the \emph{small} data set used in section~\ref{sec:reconstruction}. 
A flat universe with a cosmological
constant $\Lambda$ and $\Omega_{M}=0.3$ was used as fiducial cosmology. 
The parameters $A_{1}$ and $A_{2}$ were fitted
to each simulated data set with a fixed value of $\Omega_{M}$.

\begin{figure}
\begin{center}
\resizebox{1.0\textwidth}{!}{\includegraphics{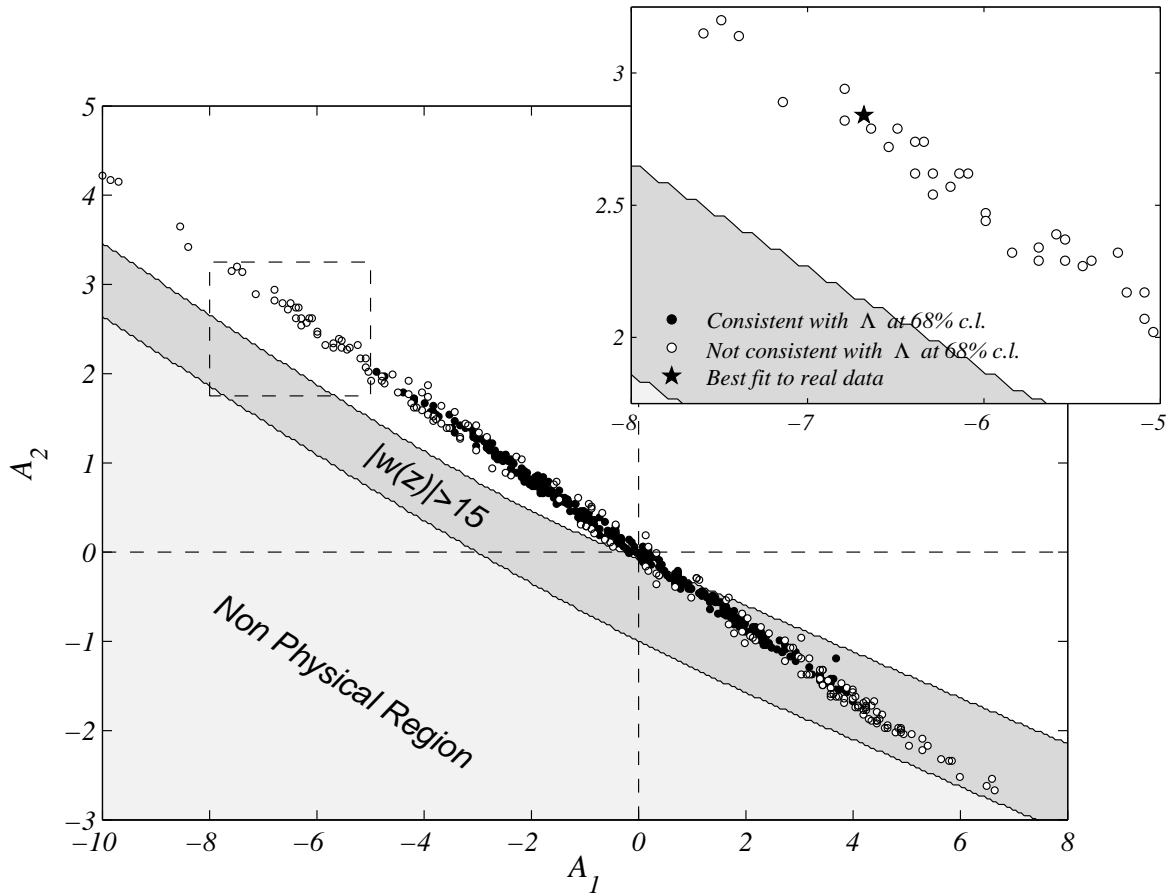}}
\caption{\label{fig:A1A2sim} A scatter plot of the best fit central
  values of $A_{1}$ and $A_{2}$ to
  $500$ simulated data sets in a universe with a cosmological
  constant. The used cosmology corresponds to
  $A_{1}=A_{2}=0$. The best fit values $A_{1}$ and $A_{2}$ are evidently
  highly correlated. The plot in the upper right--hand corner is an
  enlargement of the part inside the dashed rectangle. The filled
  circles represent simulated data sets consistent at the $68\%$
  confidence level with
  the fiducial cosmology. Data sets not consistent on this level are
  indicated by open circles. The star indicates the best fit value to
  real data. 
  The dark
  energy equation of state reconstructed from data sets
  indicated by open circles in the neighborhood of the star, would thus
  look very similar to what we find from real data~(see
  figure~\ref{fig:hzt_w}).
  Examples of $w_{DE}(z)$  reconstructed from a few of these data sets
  are shown in figure~\ref{fig:exmpl_sim}.}
\end{center}
\end{figure}

\begin{figure}
\begin{center}
\resizebox{1.0\textwidth}{!}{\includegraphics{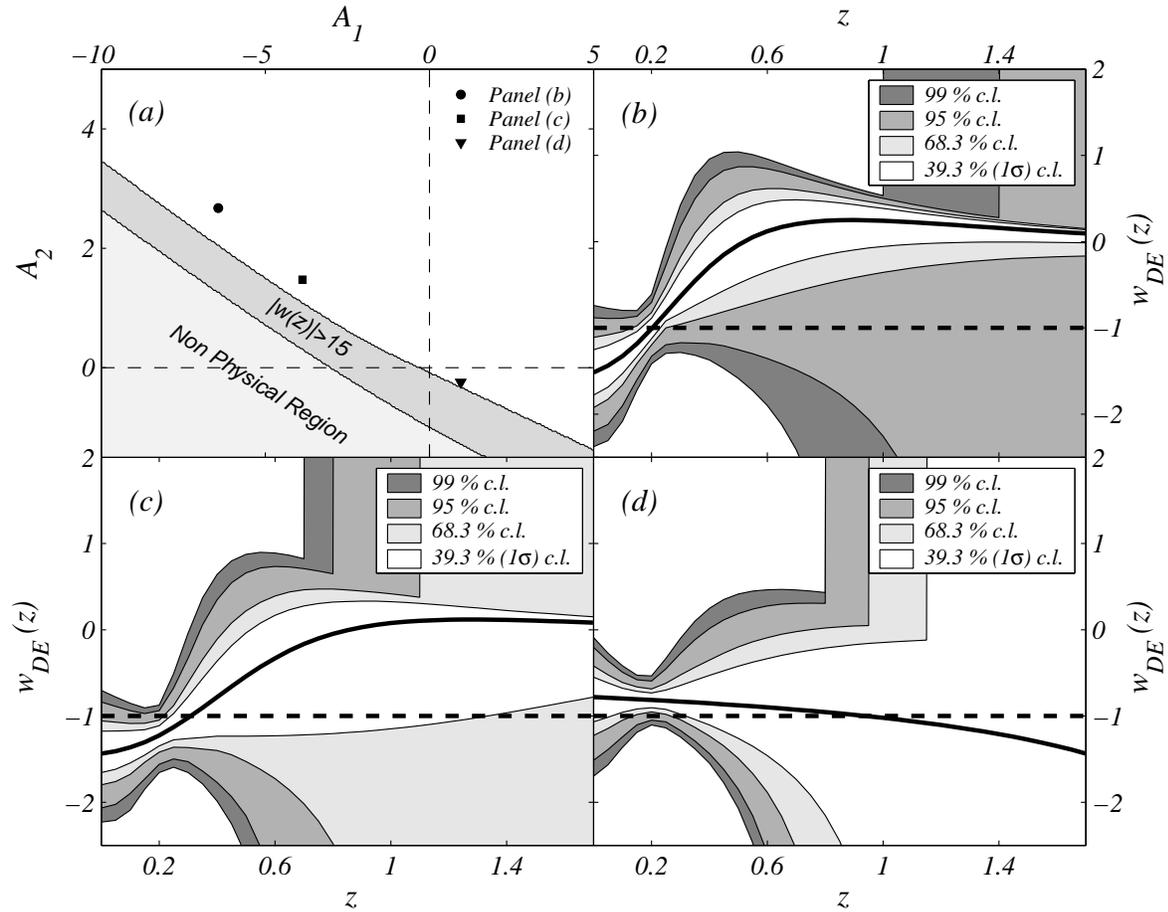}}
\caption{\label{fig:exmpl_sim} Examples of $w_{DE}(z)$ reconstructed  
  from three different simulated data sets with a cosmological constant
  are depicted in panel (b), (c) and (d). The distributions of
  redshifts and magnitude errors for these data sets were the same as
  for the \emph{small} data set used in  section~\ref{sec:reconstruction}.
  Best fit values
  to the three data sets are indicated in panel (a). The reconstructed
  equation of state parameter evolves rapidly in all three cases, which
  indicates that metamorphosis is common also in a $\Lambda$--universe.}
\end{center}
\end{figure}

A scatter plot of best fit values to the simulated data sets is shown
in figure~\ref{fig:A1A2sim}. From this plot it is clear that the best
fit values of the parameters $A_{1}$ and $A_{2}$ are highly correlated,
as anticipated in section~\ref{sec:force}.
Of the simulated
data sets, $61.8\%$ were found to be consistent with the cosmological constant
at the $68\%$ CL. 
These are indicated by full circles
in the plot. Data sets not consistent with the cosmological
constant at the $68\%$ CL, on the other hand,  are
indicated by open circles. 
The best fit to real data obtained in
section~\ref{sec:reconstruction} is indicated with a star in the expanded
figure in figure~\ref{fig:A1A2sim}. The expansion
shows a region where simulated data sets have a reconstructed
dark energy equation of state parameter similar to the one obtained from real
data. The simulated data sets in this region are inconsistent with
the fiducial cosmology at the $68\%$ CL. 
Figure~\ref{fig:exmpl_sim} shows a few examples of $w_{DE}(z)$
reconstructed from simulated data sets. The reconstructed $w_{DE}(z)$
in panel (b) is almost
indistinguishable from the one obtained from real data
(see figure~\ref{fig:hzt_w}). 
Rapidly evolving equation of state parameters, resembling the
behaviour of the result obtained from real data, can thus be obtained
from simulated data sets in a $\Lambda$
cosmology ($w_{DE}=-1$)  with the same
distributions of redshifts and magnitude errors as the real set.

\begin{figure}
\begin{center}
\resizebox{1.0\textwidth}{!}{\includegraphics{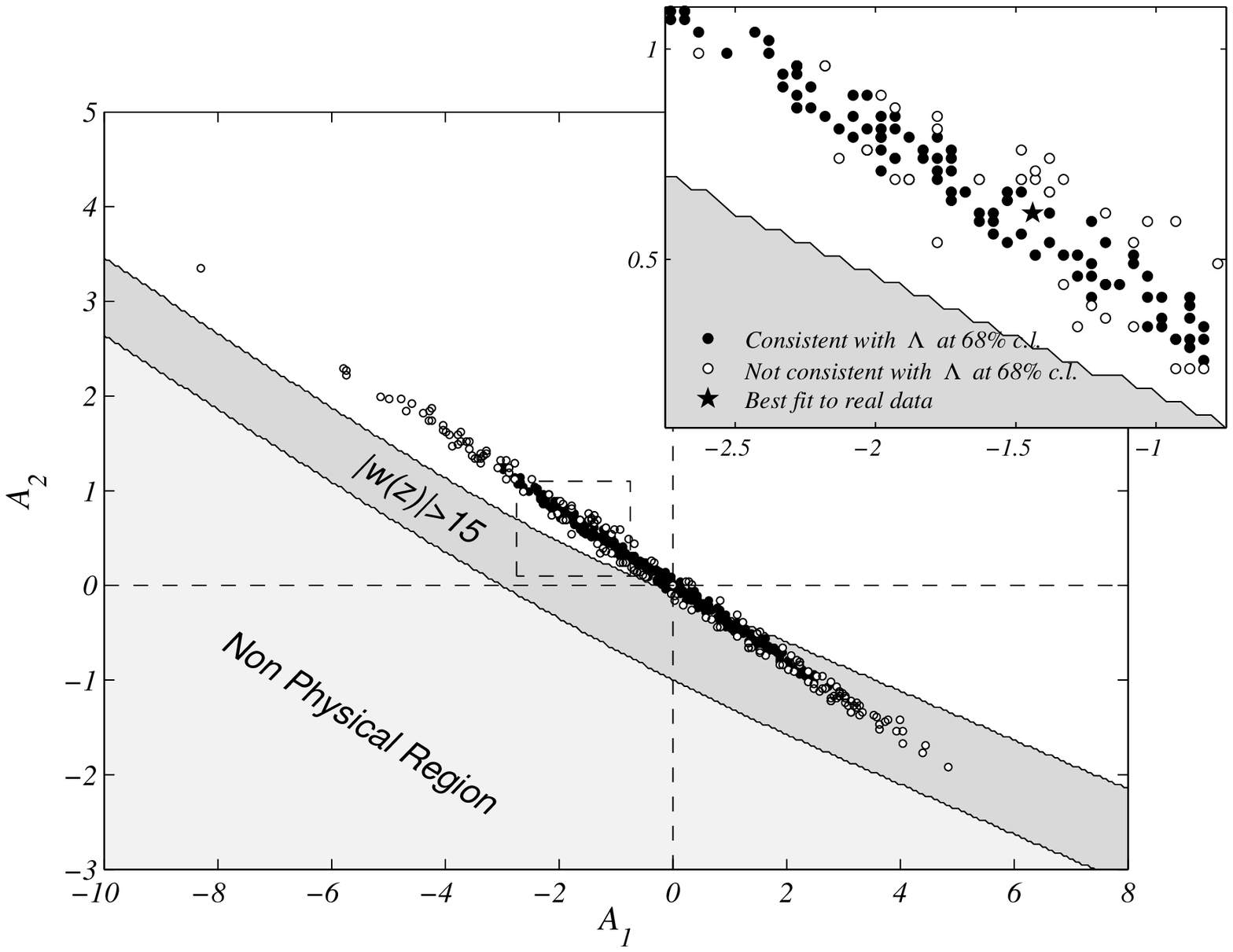}}
\caption{\label{fig:A1A2sim_16hz} The same as in
  figure~\ref{fig:A1A2sim} but the simulated data sets contain $16$
  additional supernovae at high redshifts. In this figure the star in
  the expanded figure indicates the best fit to the extended real data set.
  The additional supernovae
  have reduced the scatter of the best fit central values to the parameters
  $A_{1}$ and $A_{2}$.}
\end{center}
\end{figure}

We also simulated $500$ data sets with the same distribution of
redshifts and magnitude
errors as the \emph{large} data set used in section~\ref{sec:reconstruction}.
Figure~\ref{fig:A1A2sim_16hz}
shows a scatter plot of the best fit values to these simulations. In
this case $64.9\%$ of the simulated data sets were consistent with
the fiducial cosmology on the $68\%$ CL. As can be seen from 
figure~\ref{fig:A1A2sim_16hz}, the $16$ additional high redshift
supernovae decrease the spread of the best fit values. 

To save computational time, a simplified fitting procedure was used in
the reconstruction of $w_{DE}(z)$ from the simulated data sets, as
$\Omega_M$ was kept fixed at the input value. The value of the matter
density affects the slope of the confidence contour. By using a more
realistic fitting procedure the spread of the best fit values would increase.


\section{Extended ansatz}
As we have seen above, equation~(\ref{eq:de_eos}) involves some difficulties 
associated with the reconstruction of the dark
energy equation of state parameter. Could an extension of the ansatz overcome 
these difficulties? Studies of an arbitrary power series
describing the dark energy density might
help us to answer this question
\begin{equation}
\label{eq:power}
\rho_{DE}= \sum_{k}A_{k}x^{k},
\end{equation}
where $k$ can be zero or any positive or negative integer. 
Perturbations in the parameters $A_{k}$ will cause a change in 
$\rho_{DE}$ that is independent of these parameters 
\begin{equation}
\delta \! \rho_{DE} \approx \sum_{k}x^{k} \delta \! A_{k}.
\end{equation} 
A description of the density of dark energy in terms of a power series
will thus not favour any particular model. The equation of state
generated by a power series is given by~(\ref{eq:w(H)})
\begin{equation}
\label{eq:eos_power}
w_{DE}(x)=-\frac{\sum_{k}\frac{3-k}{3}A_{k}x^{k}}{\sum_{k}A_{k}x^{k}}.
\end{equation}
If only one coefficient in~(\ref{eq:eos_power}) is non--zero,
$w_{DE}(x)$ assumes a constant value. The power
series parametrization of the equation of state parameter is hence
capable of reproducing a whole spectrum of
constant $w_{DE}(x)$ separated by $\Delta w=\case13$
\begin{equation}
w_{DE}(x,A_{j} \neq 0,A_{k \neq j}=0)=-1+\frac{j}{3}.
\end{equation}
The cosmological constant corresponds to $j=0$. The high redshift
limit for a truncated power series with leading term $x^{n}$ resembles
this spectrum
\begin{equation}
\lim_{x \rightarrow \infty} w_{DE}(x)=-1+\frac{n}{3}.
\end{equation}
The change in $w_{DE}$
due to any perturbation of the parameters $A_{k}$ in the neighborhood
of the cosmological constant ($A_{0} \neq 0$ and $A_{k \neq 0}=0$) is
given by
\begin{equation}
\delta \! w_{DE} \approx \sum_{k} \frac{k}{3A_{0}}x^{k} \delta \! A_{k}.
\end{equation} 
The effect of perturbations in the parameters depends upon 
the leading term $A_{n}x^{n}$ in the power series. If $n$ is positive,
the change in the equation of state parameter will increase with redshift.
However, if $n$ is negative the change will decrease with redshift.
In order to discriminate between the cosmological constant
and evolving DE alternatives, we have to contrast this
behaviour with the curves describing evolution. The perturbed equation
of state parameter corresponding to non--zero $A_{k}$ at a high
redshift ($x \gg 1$) with the leading term $A_{n}x^{n}$ is:
\begin{equation}
\delta \! w_{DE} \approx -\sum_{k}^{n-1} \frac{(n-k)}{3A_{n}}x^{-(n-k)}
\delta \! A_{k}
+\frac{A_{n-1}}{3A_{n}^{2}}x^{-1} \delta \! A_{n}.
\end{equation} 
The effects of perturbations decrease with redshift if $n$ is
positive, or increase with redshift if $n$ is negative. 
This is the very opposite of the behaviour for the case of a
cosmological constant. 
The instability problem discussed above can
thus not be resolved by adding extra terms to the ansatz~(\ref{eq:ansatz}).

\section{Discussion}
The ansatz proposed by Alam \etal may be useful for modelling the dark
energy density, but its usefulness for revealing the nature of the
DE seems limited. The parametrization of the dark energy equation of
state parameter 
based on a truncated Taylor series involves a number of severe problems.  
Evolution is both favoured and forced by this parametrization. The
cosmological constant is thus mistreated by the ansatz proposed
in~\cite{alam:dec03, alam:mar04}. 
Not even an extension of the ansatz seems to be able to
overcome these difficulties.

The equation of state parameter expressed as in
equation~(\ref{eq:de_eos}) also diverges in large regions of the parameter
space. More data, which focus the solutions to the stable regions, may
solve this problem. However, the region 
close to the point describing a universe with a cosmological constant
will be in a disfavoured part of the parameter space.

The dark energy equation of state parameter reconstructed from the
data sets presented in reference~\cite{hzt:tonry}
and~\cite{hzt:barris} is inconsistent with the cosmological constant
at the $68\%$ confidence level. Simulations show that the rapidly
changing behaviour of $w_{DE}(z)$ could also
be expected with this ansatz for a $\Lambda$ universe. The
scatter of the best fit parameters  can be
reduced if additional data points at high redshift are added to the
simulated data sets. Our best fit to real data with $16$ additional
high redshift supernovae was consistent with the cosmological constant
at the $68\%$ confidence level.

We concluded that the suggested ``model independent'' method of
reconstructing the dark energy equation of state parameter in 
\cite{alam:dec03}, is in fact model 
dependent, and that the seemingly striking results are  
likely to be due to this deficiency.

\ack
A.G. is a Royal Swedish Academy
Research Fellow supported by a grant from the Knut and Alice
Wallenberg Foundation.

\end{document}